\documentclass[preprint,pre]{revtex4}

\usepackage{graphicx}
\usepackage{amssymb}

\begin{document}

\title{ Experimental test of the Gallavotti-Cohen fluctuation theorem in
turbulent flows}

\author{S. Ciliberto}
\author{N. Garnier}
\author{S. Hernandez}
\altaffiliation{Present address: UNAM, Dept. Sciencias, Mexico}
\author{C. Lacpatia}
\author{J.-F. Pinton}
\author{G. Ruiz Chavarria}
\altaffiliation{Present address: UNAM, Dept. Sciencias, Mexico}
\affiliation {Laboratoire de Physique, C.N.R.S. UMR5672, Ecole
Normale Sup\'erieure de Lyon,  46,  All\'ee d'Italie,  69364 Lyon,
France}

\date{\today}
\centerline{submitted to Physica A}

\begin{abstract}
We test the fluctuation theorem from measurements in turbulent
flows. We study the time fluctuations of the force acting on an
obstacle, and we consider two experimental situations: the case of
a von K\'arm\'an swirling flow between counter-rotating disks (VK)
and the case of a wind tunnel jet. We first study the symmetries
implied by the Gallavotti-Cohen fluctuation theorem (FT) on the
probability density distributions of the force fluctuations; we
then test the Sinai scaling. We observe that in both
experiments the symmetries implied by the FT are well verified,
whereas the Sinai scaling is established, as expected, only for
long times.
\end{abstract}

\maketitle

\section{Introduction}
The fluctuations of global quantities in out-of-equilibrium systems is a subject
of current interest which presents many unsolved problems. One of these problems
is the prediction of the shape of the probability density function (PDF) of the
variable under study; the PDF is of course not necessarily Gaussian~\cite{Holdsworth}.
Among several approaches used to study these problems, the theory of Gallavotti
and Cohen~\cite{galla1,galla2,galla3,Kurchan,Ramses}, based on a chaotic
hypothesis, leads to interesting predictions. The main prediction of the Gallavotti-
Cohen fluctuation theorem (FT)~\cite{galla1} concerns the probability density
function of a variable related to  the phase space contraction rate of the
nonequilibrium system under study, this variable being a current $j$ such as
the flux of heat, or of momentum, or energy, etc. The mean of $j$ on a time
interval $\tau$ is defined as
\begin{equation}
J_\tau= {1 \over \tau} \int_t^{t+\tau} j(t') dt' \ .
\end{equation}
One is then interested in the  the probability distribution
$\pi_\tau(Y)$ of the variable $Y=J_\tau/J_{\infty}$,  where
$J_{\infty}=\lim_{\tau \rightarrow \infty} J_\tau$ is the
stationary average current studied in the system.  If $\tau $ is larger
than a characteristic time of the system, then the chaotic
hypothesis~\cite{galla1} predicts that $\pi_\tau(Y)$ verifies
\begin{equation}
\ln { \pi_\tau(Y) \over \pi_\tau(-Y)}= \tau \ \sigma \ Y
\label{FT}
\end{equation}
where $\sigma$ is proportional to the phase space contraction rate. It
is important to stress that the above equation is valid for all values
of $Y$. A related expression for the probability density $\pi_\tau(Y)$
has been established by Sinai under quite general assumptions:
$\pi_\tau(Y)=A_\tau \ e^{-\zeta(Y) \ \tau}$. In this formulation the
function $\zeta(Y)$ is a limiting function, for $\tau \rightarrow
\infty$,
\begin{equation}
\zeta(Y) ={ \ln [\pi_\tau(Y)]- \ln[ A_\tau] \over \tau} \ ,
\label{Sinaifunction}
\end{equation}
independent of $\tau$~\cite{Sinai}. However, the FT (with the addition
of the chaotic hypothesis) is giving more information: the function
$(\zeta(Y)-\zeta(-Y))$ is linear in $Y$ and its derivative is related to
the contraction rate in phase space. In this sense, the FT takes into
account the underlying dynamics taking place in the system.

However, in the original derivation of the fluctuation theorem, the
proof requires several very restrictive hypothesis, among which the time
reversibility of the system. Therefore it is important to check the
symmetry predicted by eq.\ref{FT} in more realistic systems in order to
probe the universality and applicability in practical cases. Eq.\ref{FT}
has been tested with success on a numerical simulation of a rather
artificial system~\cite{galla4}, and it is also quite well verified for
the heat flux in a chain of coupled non-linear oscillators~\cite{Livi,
Aumaitre} and in the energy dissipation rate in the shell model of
turbulence~\cite{Aumaitre}.

The situation is less clear for local variables.
Gallavotti~\cite{galla3} suggested that the predictions of the FT could
be extended to ``the average of local observables". This has prompted us
to check this hypothesis using local measurements in a turbulent flow.
We started with preliminary tests using the data of turbulent
Rayleigh-Benard convection~\cite{Cili1}. These tests did show that the
fluctuations of a quantity proportional to the heat flux verify both
eq.\ref{FT} and eq.\ref{Sinaifunction}. In this article we consider the
case of the pressure fluctuations in a turbulent wind. Specifically, we
consider the statistics of the force applied on an object. We find that
its fluctuations verify the Gallavotti-Cohen predictions.

\section{ Experimental apparatus}
\begin{figure}[htb]
\centerline{\includegraphics[width=8cm]{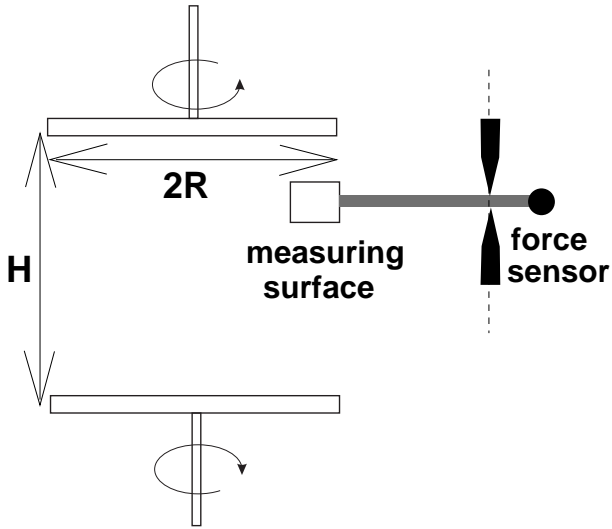}}
\caption{von K\'arm\'an experimental set-up. The discs with radius $2R=20$~cm
are set $H=20$~cm apart. They are counter rotated at the fixed frequency $\Omega
= 40$~Hz.}
\label{expset}
\end{figure}

The experimental set-up which is shown in fig.\ref{expset} consists of two
coaxial, counter rotating disks. This experimental set-up, known as the von
K\'arm\'an geometry, produces an intense turbulence in a compact region of
space. It has been previously used to study the fluctuations of the power
injected to maintain the turbulent flow at a fixed integral Reynolds number ---
more details on the experimental apparatus can be found in ref.\cite{PintonI}.

The distance between the disks is 20~cm.  As illustrated in
fig.\ref{expset}  a square plate, of side 4~cm and 1~mm thick, is
inserted between the two disks at a distance of 6~cm from one of
the two disks, with its center at 6~cm from the disks rotation
axis. The plate is mounted on an extreme of a 10~cm long lever. The
other extreme of the lever is fixed on a strain gage (ENTRAN
ELFS-T3M-10N force transducer). The lever axis, which is at
distance of 1~cm from the strain gage, is mounted on  very precise
ball bearings in order to reduce friction. With this system the
minimum detectable force exerted on the plate is 1~mN, with a
response time of $10^{-2}$~s.

The quantity under measurement is the integral of the pressure over the area
$S$ of the object:
\begin{equation}
\mathbf{F}(t) = \int_{S} \, ds \; p(\mathbf{r}, t) \ .
\end{equation}
The pressure itself $p(\mathbf{r}, t) $ is the solution of the Poisson equation
\begin{equation}
\Delta p(\mathbf{r}, t) = 2 \rho \partial_i u_j(\mathbf{r}, t) \partial_j
u_i(\mathbf{r}, t) \ ,
\label{eqPoisson}
\end{equation}
where $\rho$ is the fluid density and $\partial_i u_j$ the velocity gradient
tensor. It thus instantaneoulsy relates to the strain and vorticity fluctuations
in the flow. Note that the presence of the object introduces a no-slip condition
on its surface which changes the velocity distribution; the fluctuations of
force on the objet are related ---~but not strictly equal~--- to the flux of momentum in
the undisturbed flow in the absence of the obstacle.  The  mean value $F_0$ of
the force on the object is set by the pressure head ($p \simeq \rho U^2$), where
$U$ is the mean velocity at the location of the object. The time fluctuations of
the force are related to the small scale turbulent fluctuations, as indicated by
equation~(\ref{eqPoisson}) and shown in several studies on the fluctuations of
the local pressure~\cite{FauvePression,CadotPression}.

We analyze a set of data recorded at a disk rotation frequency of 40~Hz,
corresponding to a Reynolds number of about $Re = 2\ 10^5$. The signal
of the instantaneous force $F(t)$ is filtered at 70~Hz and sampled at
160~Hz with 21bits resolution Agilent 1430 digitizer ($\tau_0 = 1/160 =
6.25$~ms is the sampling time). The signal is recorded for about $2 \
10^4$ integral times. In these conditions the mean force $F_0$ on the
plate is $0.12$~N.

\begin{figure}[h]
\centerline{\includegraphics[width=16cm]{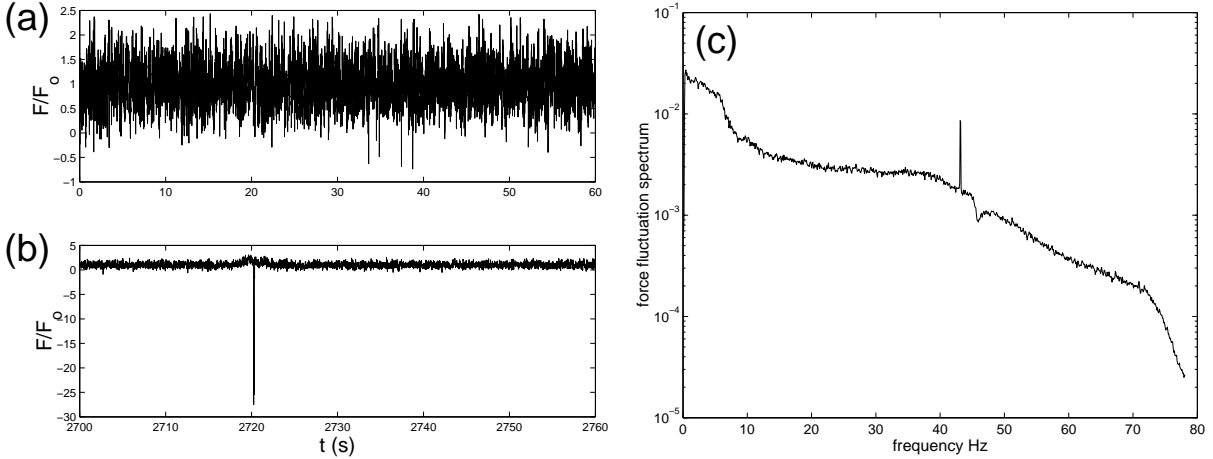}}
\caption{(a) Force of the wind on the measuring surface as  a
function of time.  (b) Expanded view of the force signal --- note the strong
negative peaks. (c) Power spectrum of the force fluctuations.}
\label{dataKarman}
\end{figure}

The variable of interest is $y=F(t)/F_0$. Its typical time evolution is
plotted in fig.\ref{dataKarman}; in (a) is shown the `standard'
fluctuations of the signal in a large time-window while in (b) we have
selected a time interval during which a negative fluctuation of large
amplitude occurs. One observes that $y=F(t)/F_0$ fluctuates mainly about
the mean value $1$, but also that there are strong negative events:
sometimes the plate is pushed in the opposite direction of the mean wind
velocity. The spectrum, plotted in fig.\ref{dataKarman}(c), is very
broad. The thin peak at 40~Hz corresponds to the disk rotation
frequency. The cut-off at 70~Hz is imposed by the filter of the
digitizer. Note on the spectrum that the change of slope near 35~Hz is
at a frequency of the order of $U/L$ where $U$ is the mean velocity at
the plate location and $L$ is the plate size.

\section{Data analysis}

\begin{figure}[h]
\centerline{\includegraphics[width=12cm]{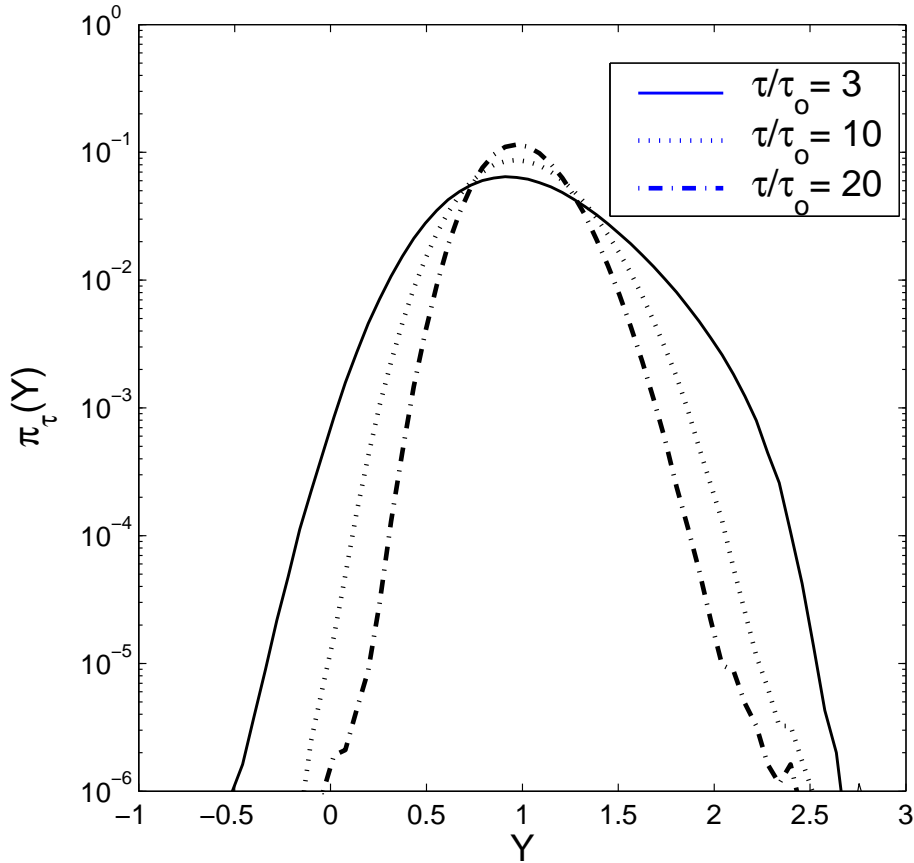}}
\caption{ Probability density function for the fluctuations of the force on an
obstacle placed in the von K\'arm\'an flow, corresponding to the time signals
shown in fig.\ref{dataKarman}(a,b)}
\label{PDF}
\end{figure}

The data analysis of the fluctuations of the normalized force $y=F/F_0$
has been performed in the following way. We first compute the integrated
quantity:
\begin{equation}
Y(t,\tau)= {1 \over \tau} \int_t^{t+\tau} y(t') dt' \,,
\end{equation}
then we calculate the histograms of the $Y(t,\tau)$ values in order to
obtain the probability density functions $\pi_\tau(Y)$ for each value of
the coarse graining time $\tau$. The logarithm of the functions
$\pi_\tau(Y)$ are plotted in fig.\ref{PDF} as a function of $Y$ for
$\tau= 3\tau_0, 10\tau_0$ and $20\tau_0$. One clearly sees that the PDFs
change with the integration time $\tau$. For all values of $\tau$, they
are non-Gaussian eventhough the force results from a spatial integration
of the pressure fluctuations. The central limit theorem does not
necessarily apply here~\cite{BHP} because the pressure fluctuations can
be correlated over the plate size: they are caused by turbulent
structures of sizes ranging from the integral to the dissipative Kolmogorov
scale. In addition, one observes the occurence of negative fluctuations,
as shown in fig.\ref{dataKarman}.

\begin{figure}[h]
\centerline{\includegraphics[width=16cm]{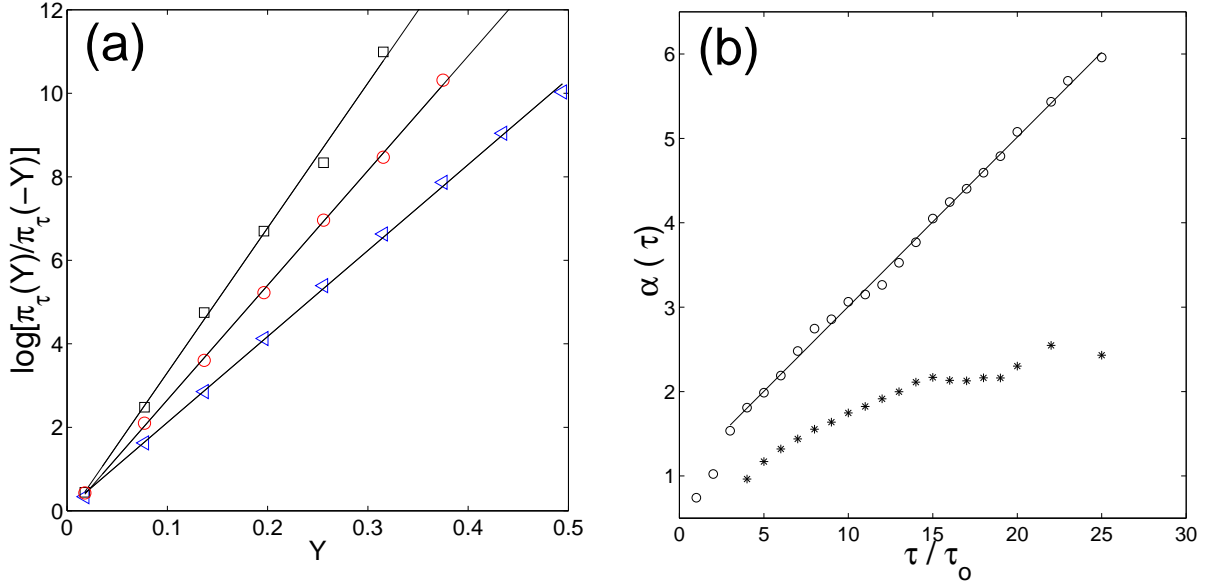}}
\caption{ (a) The function $G_\tau(Y)$, as defined in the text, plotted as a
function of $Y$; integration times: ($\triangle$) $3\tau_0$, ($\circ$)
$5\tau_0$, ($\square$) $10\tau_0$. If eq.\ref{FT} is verified then $G_\tau(Y)$
has to be a linear function of $Y$. (b) The function $\alpha(\tau)$ computed
using the values $\pi_\tau(Y)$ for  $-0.5<Y<0.5$ ($\circ$) and $1.5<Y<2.5$
($\ast$), plotted as a function of $\tau$.
}
\label{GYvY}
\end{figure}

From the PDF $\pi_\tau(Y)$, we construct the function
$G_\tau(Y)=\ln({\pi_\tau(Y)\over\pi_\tau(-Y)})$ for $0 \le Y \le 0.5$.
This small interval around zero allows us to study the symmetry of
$G_\tau(Y)$ even for the largest integration times, {\it e.g.}
$\tau=20\tau_0$. The function $G_\tau(Y)$ is plotted as a function of
$Y$ in fig.\ref{GYvY}(a) for $\tau= 3\tau_0, 5\tau_0$ and $10\tau_0$. We
find that $G_\tau(Y)$ is very much linear in $Y$, as predicted by eq.1
with a slope $\alpha(\tau)$ which increases with $\tau$.

We thus turn to the study of the slope $\alpha(\tau)$ as a
function of  $\tau$. If eq.1 is correct $\alpha(\tau)$ should be a linear
function of $\tau$. The slopes $\alpha(\tau) \;  {\rm vs.} \; \tau$ are plotted
in fig.\ref{GYvY}(b). We see that for $\tau/\tau_0 > 4$, the dependence of
$\alpha(\tau)$ is linear in $\tau$. It is a quite noteworthy feature that
$\tau/\tau_0=4$ is very close to the rotation period of the disks (this also
corresponds to the correlation time of the von Karman flow~\cite{BHP}).  Indeed,
fig.\ref{GYvY}(b) shows that the linear behaviour of $\alpha(\tau)$, predicted
by eq.1, is obtained
only for $\tau$ larger than the mean integral time of the system. This
observation will be discussed in more details in the next sections.

Here we want to stress another aspect of the measurement, namely the
importance of considering the points around $Y=0$ for the study of the
Fluctuation Theorem. As a test, we have computed the evolution
$\alpha(\tau)$ obtained by the same analysis of the symmetries of the
PDF around the point $Y=2$ instead of the point $Y=0$. Corresponding
results are plotted with $\ast$ symbols in fig.\ref{GYvY}(b). We observe
that in such a case $\alpha(\tau)$ is not linear in $\tau$. We thus find
that the point $Y=0$ has some special property; this is in agreement
with the FT prediction.

Summarizing we find that for $\tau/\tau_0 > 4$ $\alpha(\tau)$  is linear, and we
can write $\alpha(\tau)= A \tau + B$. Thus
\begin{equation}
 \ln { \pi_\tau(Y) \over \pi_\tau (-Y)} = (A \tau + B) \
Y   \label{result}
\end{equation}
We stress that this expression is not in contradiction with eq.1 which
has been derived for $\tau \gg \tau_0$; rather eq.\ref{result} should be
considered as a non- asymptotic version of eq.1.

\begin{figure}
\centerline{\includegraphics[width=12cm]{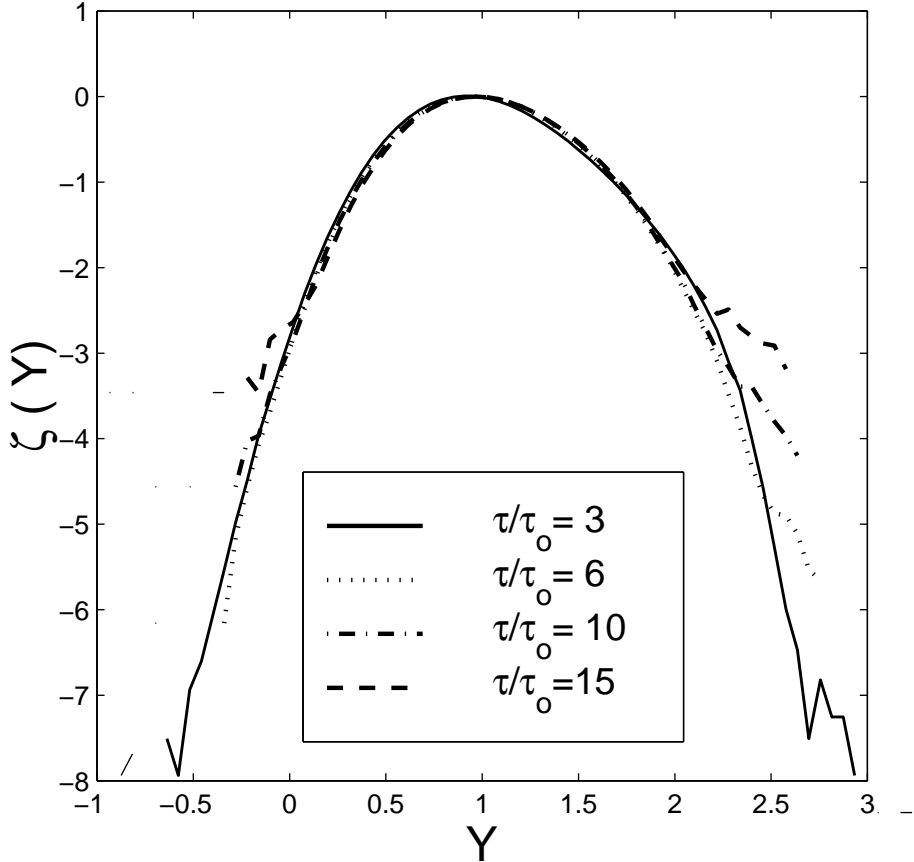}}
\caption{Sinai scaling function obtained from the PDF measured in the counter
rotating disk experiment }
\label{zetavy}
\end{figure}

Finally, we check the existence, for the variable of our experiment, of
a function independent of $\tau$ similar to the one defined in eq.3. In
order to do this test, we write the PDF as $\pi_\tau(Y)=A_\tau \
e^{-\zeta(Y) \ \alpha(\tau)}$. Here we use the previously determined
function $\alpha(\tau)$ instead of $\tau$, because, as discussed for
eq.2, we are not in the asymptotic regime for all values of $\tau$. If
the prediction of ref.\cite{Sinai} can be applied in this case, then
$\zeta(Y)=\ln[P(Y)/A_\tau]/\alpha(\tau)$ has to be independent of
$\tau$. The functions obtained from the PDFs $\pi_\tau(Y)$ of
fig.\ref{PDF} is plotted in fig.\ref{zetavy}. We clearly see that the
rescaling is very good. In particular we note that the positive part of
$\pi_\tau(Y)$, which is strongly non-Gaussian, also scales perfectly.
Thus the $\alpha(\tau)$ determined using eq.3 allows us to rescale the
PDF of $Y$ for any $\tau$. It is important to stress that the rescaling
is not possible if $\tau$, instead of $\alpha(\tau)$, is used.
Furthermore it is obvious that no scaling can be observed using the
$\alpha(\tau)$ obtained by performing the analysis around the point
$Y=2$.

In order to check the Sinai scaling function  we have performed a
different data analysis as suggested in ref.~\cite{Evans}. This
analysis, which we will not describe here, yields results in close
agreement with those plotted in fig.\ref{zetavy}.

\section{ A wind tunnel experiment}

In the previous section we have shown that the PDF of the fluctuations
of force on a small surface satisfies the symmetries imposed by FT and
that they can be rescaled using eq.\ref{Sinaifunction}. These two
properties were also observed on the PDFs of the local heat flux
fluctuations in a turbulent thermal convection experiment~\cite{Cili1}.
It is important to recall that eq.\ref{FT} has been proved for global
observables and the fact that the symmetry imposed by eq.\ref{FT} can be
observed on a local measurement is not obvious. In addition, the counter
rotating disks experiment and the thermal convection experiment are
spatially bounded flows. Indeed, in the von K\'arm\'an experiment, one
observes that, eventhough the fluid is not confined within walls, the
turbulence remains bounded in a volume whose lateral extent is roughly
equal to twice the diameter of the driving disks. In such situations,
one can argue that a local measurement is indicative of the global
behavior of the system.

\begin{figure}[hb]
\centerline{\includegraphics[width=15cm]{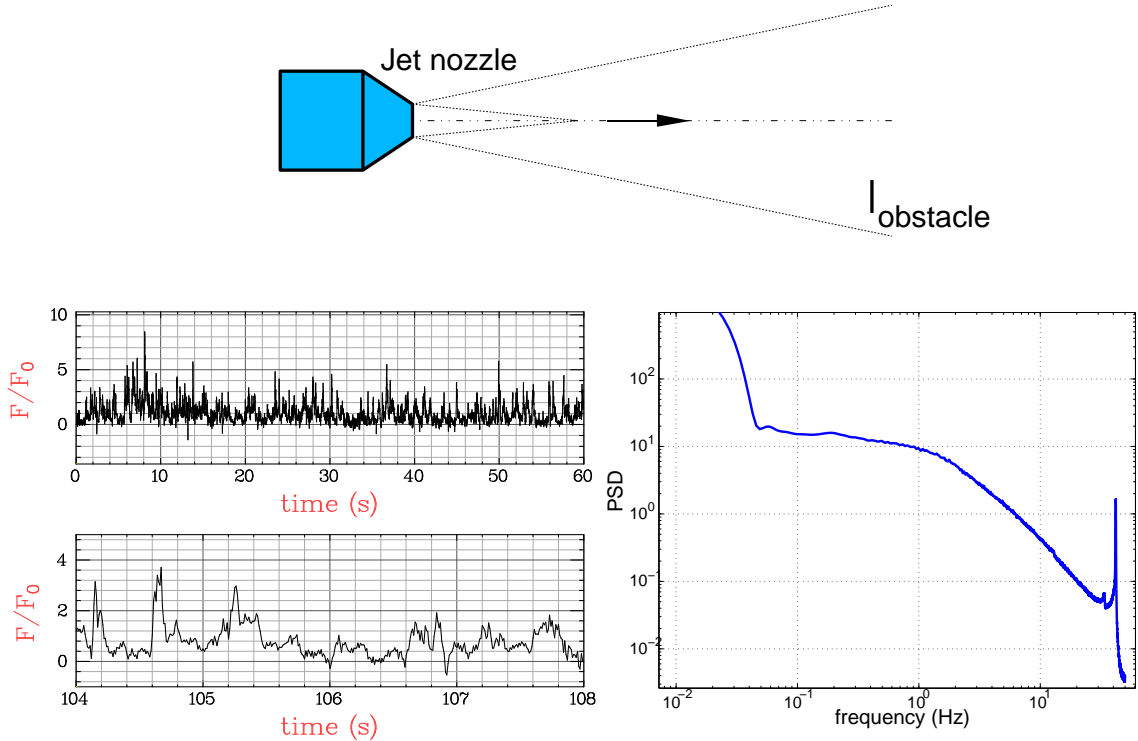}} \caption{ (a)
Schematic diagram of the wind tunnel experiment. (b) and (c)
Typical time series of the force fluctuations at $Re=45000$. (d)
Spectrum of the force fluctuations measured in the  wind tunnel at
$Re=45000$.  } \label{PSDtunnel}
\end{figure}

The question thus arises naturally as whether the same observations can
be made in truly open systems, where velocity fluctuations are advected
outside the observation zone. For this reason, we have repeated the
experiment on the wind pressure fluctuations inserting the measuring
surface in the turbulent region of a jet developing inside a wind
tunnel. The jet is formed by a 5.2~cm nozzle. The obstacle is located 35
diameters downstream of the nozzle where the turbulence is fully
developed, at distance of 4 diameters off the axis of the jet. At the
measurement location, the mean wind speed is 1~m/s and the Reynolds
number based on the jet size is $ Re=45000$. The force of the wind on the
obstacle is measured as the bending of the shaft that holds the plate,
as detected from the deviation of a laser beam. The sampling frequency
is $1/\tau_0 = 100$~Hz, and $3.6 \ 10^6$ data points are accumulated, with
a 16~bit resolution.

\begin{figure}[t!]
\centerline{\includegraphics[width=10cm]{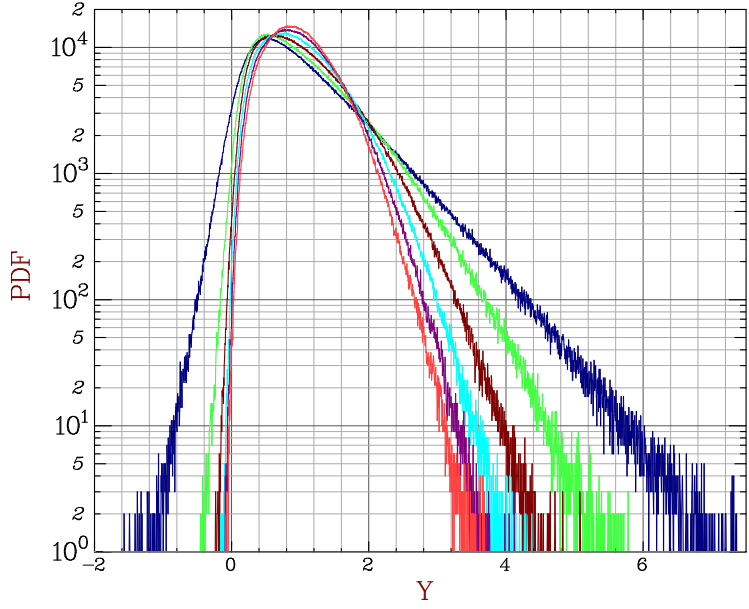}} \caption{ Wind
tunnel experiment. PDF of $Y$ for $\tau/\tau_0=1, 10, 20, 30, 40,
50$, upper to lower curve.}
\label{PDFtunnel}
\end{figure}

The force fluctuation spectrum is plotted in fig.\ref{PSDtunnel}(b). The
spectrum is very broad and presents no resonance (the peak at 30Hz is
due to the wind tunnel driving fan). As in sec.3 we define the
normalized variable $y=F(t)/F_0$, and the coarse-grained variable $Y$
after integration over a time interval $\tau$. The PDFs of $Y$ are
plotted in fig.\ref{PDFtunnel}. The curves are extremely non-Gaussian,
with the development of an exponential tail at large values of the
force, but also important negative excursion on the left-hand side.

\begin{figure}
\centerline{\includegraphics[width=16cm]{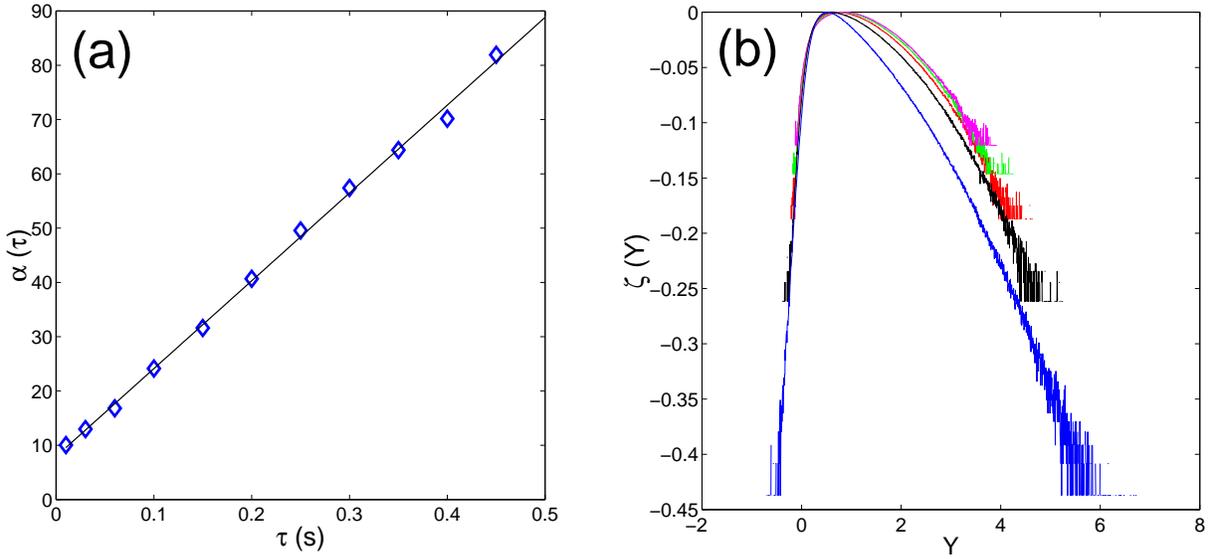}}
\caption{ Wind tunnel experiment. (a) $\alpha(\tau)$ and (b) $\zeta(Y)$,
estimated as in sec.3}
\label{alfatunnel}
\end{figure}

In order to analyze them, we repeat on the PDFs of the wind pressure in
the wind tunnel the analysis developed in section~3. Specifically we
verify that the previously defined function $G_\tau(Y)$ is linear in $Y$
for $-0.5<Y<0.5$. In fig.\ref{alfatunnel}(a) we plot the measured slope
$\alpha(\tau)$ as a function of $\tau$. We find that the affine behavior
previously observed, $\alpha(\tau)=A \tau +B$, is again verified here
even for small values of $\tau$. It means that the PDFs of the wind
tunnel measurements do verify the functional form of the fluctuation
theorem (eq.\ref{FT}), in an interval close to $Y=0$.

However we find that for small $\tau$ values, it is not possible to
derive a unique function $\zeta(Y)$ independent of $\tau$ which would
yield a complete rescaling of the PDFs. This is evidenced in
fig.\ref{alfatunnel}(b) where it can be observed that some rescaling
property begins to form at the larges values of $\tau$, {\it e.g.} for
$\tau > 20\tau_0$. It is very interesting to note that this time
corresponds to the time of flight of the wind from the jet outlet to the
measuring point; for times longer than this time of flight, the
fluctuations can be associated to global variations of the turbulent
field impacting the obstacle, as a block. This point deserves more
investigations, some currently under progress.

\section{Concluding remarks}

In this paper we have described two experiments in which we have
measured the pressure fluctuations of the wind on a small surface. One
of the two experiments, the counter rotating disks, is a spatially
correlated flow~\cite{BHP} and the other, the wind tunnel, is an open
flow. The main results of this investigation are that in the spatially
correlated flow the PDFs of the fluctuations of force on an obstacle,
which are strongly non Gaussian, verify the symmetry imposed by the
Gallavotti-Cohen fluctuation theorem, and in addition can be rescaled
according to the Sinai theory. More precisely in order to rescale the
PDF one has to use the measured $\alpha(\tau)$ and not simply $\tau$ as
indicated in eq.\ref{Sinaifunction}. The linear function $\alpha(\tau)$
may be seen as a finite time correction of eq.\ref{Sinaifunction}, which
is valid only for $\tau\rightarrow\infty$. The existence of a scaling
function $\zeta(Y)$ independent of $\tau$ is very important because it
strengthen the check of the FT. Indeed a strict test of the FT using
eq.\ref{FT} can only be done for values of $Y$ close to zero, and one
always observe that the probability of getting negative fluctuations in
an actual system is actually very small. However our measurements show
that using the value of $\alpha(\tau)$ estimated for $Y$ close to zero,
it is possible to scale all the PDFs onto a single function $\zeta(Y)$
for all $Y$. This is a non trivial property which was already observed
in the heat fluctuations in turbulent thermal convection~\cite{Cili1}.
This property is not observed, at least for short times, in the jet data
described in sec.4 of this paper. The main difference between the jet
experiment, the thermal convection and counter rotating disk experiments
is that the the first behaves as a spatially uncorrelated flow because
perturbations are always advected downstream of the nozzle. As we have
already mentioned this difference is an important one. Indeed
eq.\ref{FT} has been proved only for global observables, therefore one
can picture that if the turbulence is strong enough, a quasi local
measurement can provide a significant measurement of all of the flow, if
the later is spatially correlated. This is not necessary true in a open
flow where the recurrence time can be very long. This claim is of course
preliminary and it merits to be studied in more details.

Another important point which deserves a study is the prefactor in
eq.\ref{FT} which in the standard derivation is the phase space
contraction rate. In our experiments we have to study how this prefactor
changes as a function of the Reynolds number. This is needed in order to
have a more quantitative comparison between theory and experiment.

Finally we have to stress that the study of the probability of extreme
negative fluctuations can be very useful in many applications. If one
could prove the universality of eq.\ref{FT} in various geometries, then
one would readily use it to compute the probability of the rare,
extremely negative, events.

\vspace*{6mm}
{\bf Acknowledgements} We acknowledge useful discussions with E.
Cohen, G. Gallavotti and L. Rondoni.


\end{document}